\documentclass[sigconf, authorversion, balance]{acmart}

\copyrightyear{2025}
\acmYear{2025}
\setcopyright{rightsretained}
\acmConference[RecSysChallenge '25]{RecSys Challenge 2025}{September 22, 2025}{Prague, Czech Republic}
\acmBooktitle{RecSys Challenge 2025 (RecSysChallenge '25), September 22, 2025, Prague, Czech Republic}
\acmDOI{10.1145/3758126.3758131}
\acmISBN{979-8-4007-2099-4/25/09}

\begin{document}

\title[4th Place Solution in RecSys Challenge 2025]{Blending Sequential Embeddings, Graphs, and Engineered Features: 4th Place Solution in RecSys Challenge 2025}

\author{Sergei Makeev}
\email{neuralsrg@gmail.com}
\orcid{0009-0003-5451-6475}
\affiliation{%
  \institution{Yandex}
  \city{Moscow}
  \country{Russia}
}

\author{Alexandr Andreev}
\email{sasha.andrieiev.2001@mail.ru}
\orcid{0009-0002-3826-3060}
\affiliation{%
  \institution{Yandex}
  \city{Moscow}
  \country{Russia}
}

\author{Vladimir Baikalov}
\email{nonameuntitled159@gmail.com}
\orcid{0009-0009-4864-2305}
\affiliation{%
  \institution{Yandex}
  \city{Moscow}
  \country{Russia}
}

\author{Vladislav Tytskiy}
\email{vladtytskiy@gmail.com}
\orcid{0009-0003-3960-8689}
\affiliation{%
  \institution{Yandex}
  \city{Moscow}
  \country{Russia}
}

\author{Aleksei Krasilnikov}
\email{TheKabeton@yandex.ru}
\orcid{0009-0007-3484-7970}
\affiliation{%
  \institution{Yandex}
  \city{Moscow}
  \country{Russia}
}

\author{Kirill Khrylchenko}
\email{elightelol@gmail.com}
\orcid{0009-0007-3640-8795}
\affiliation{%
  \institution{Yandex}
  \city{Moscow}
  \country{Russia}
}

\renewcommand{\shortauthors}{Makeev et al.}

\begin{abstract}
  This paper describes the 4th-place solution by team \texttt{ambitious} for the RecSys Challenge 2025, organized by Synerise and ACM RecSys, which focused on universal behavioral modeling. The challenge objective was to generate user embeddings effective across six diverse downstream tasks. Our solution integrates (1) a sequential encoder to capture the temporal evolution of user interests, (2) a graph neural network to enhance generalization, (3) a deep cross network to model high-order feature interactions, and (4) performance-critical feature engineering.
\end{abstract}




\maketitle

\section{Introduction}
E-commerce recommender systems are typically tasked with predicting relevant items, churn risk, or purchase propensity. The goal of the RecSys Challenge 2025\footnote{https://www.recsyschallenge.com/2025/} was to develop a Universal Behavioral Profile~--- a user representation that generalizes across multiple such tasks. The final leaderboard is shown in Table~\ref{tab:leaderboard}.

\begin{table}
    \small
    \setlength{\tabcolsep}{3pt}
    \caption{Final leaderboard rankings. hid.1, hid.2, and hid.3 refer to three hidden tasks.}
    \label{tab:leaderboard}
    \begin{tabular}{cc|cccccc}    
        \toprule
        Rank & Team & Churn & Category & SKU & hid.1 & hid.2 & hid.3 \\
        \midrule
        1 & rec2 & 0.7375 & 0.8179 & 0.8224 & 0.7717 & 0.8293 & 0.8161 \\
        2 & ai\_lab\_recsys & 0.7376 & 0.8180 & 0.8130 & 0.7649 & 0.8052 & 0.8117 \\
        3 & SenseLab & 0.7374 & 0.8180 & 0.8081 & 0.7637 & 0.8081 & 0.8116 \\
        4 & \textbf{ambitious} & 0.7389 & 0.8151 & 0.8041 & 0.7699 & 0.7948 & 0.8147 \\
        \bottomrule
    \end{tabular}
    \Description{Table presents final leaderboard metrics for the best four teams.}
\end{table}

Participants were required to produce embeddings for one million users based on a rich set of historical activities: cart additions, purchases, cart removals, page visits, and search queries. These embeddings were then used server-side to train six downstream models: for churn prediction, purchased category prediction, and purchased item prediction; the remaining three tasks were undisclosed to participants throughout the competition.

In this paper, we describe a recipe that led to strong performance in the RecSys Challenge 2025:
\begin{itemize}
    \item We design a heterogeneous causal transformer-based sequential encoder trained in a multi-task manner.
    \item We adopt an industry-scale graph neural network TwHIN\,\cite{el2022twhin}.
    \item We share insights into feature engineering techniques that contributed substantially to performance.
    \item We develop a neural network based on DCN-v2\,\cite{wang2021dcn} that fuses handcrafted features with sequential user representations.
\end{itemize}
\section{Proposed Approach}

To obtain user embeddings, we aggregate the representations produced by three distinct models: a sequential encoder, the TwHIN graph neural network, and a neural network based on the DCN-v2 architecture. The aggregated representations are then concatenated to the standardized numeric features.
The illustration of our solution is provided in Appendix~\ref{sec::appendix_solution}.
In this section, we detail the architecture and training procedure for each component of our ensemble.

\subsection{Sequential Encoder}
\label{sec:sequential}

\begin{figure*}[h]
    \centering
    \includegraphics[width=0.9\textwidth]{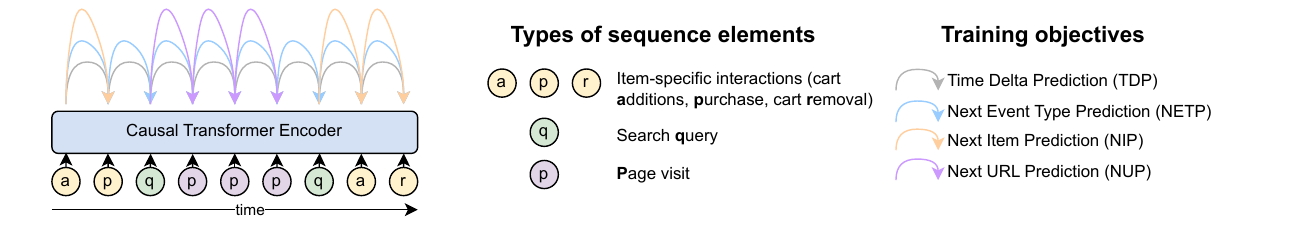}
    \caption{The sequential encoder processes heterogeneous user sequences. For every position, the model predicts the time delta and event type of the next sequence element. Hidden states preceding item-specific and page visit events are used for Next Item Prediction and Next URL Prediction tasks, respectively.}
    \Description{Figure depicts a unified user sequence consisting of five types of events. Each Transformer output contributes to the described prediction losses.}
    \label{fig:seq_enc}
\end{figure*}

\subsubsection{Model Inputs}
We merge all available user events (cart additions, purchases, cart removals, page visits, and search queries) into a single heterogeneous chronological sequence. During training, user histories exceeding \(N\) events are split into chunks of \(N\) consecutive events. For inference, we use the \(N\) most recent events to compute the sequential user embedding.

We divide event features into content and context features. Event embedding is constructed as the sum of its content and context embeddings.

Content features include:
\begin{enumerate}
    \item In the provided data, prices are quantized into 100 discrete bins. We encode them using ID-based embeddings, with each embedding corresponding to a price bin.
    \item Textual features (i.e., item names and queries) are represented by LLM-based embeddings of length 16, with each dimension already quantized into one of 256 bins. We use a shared embedding table for both names and queries to map each bin to a vector. The final representation is obtained using an embedding bag that averages the 16 resulting vectors.
    \item There are high-cardinality categorical features: SKU, category, and visited URL IDs. We encode them with Multi-size Unified Embedding framework\,\cite{coleman2023unified}. We use a shared embedding table with a vocabulary size of 524288 and perform two lookups per feature value.
\end{enumerate}

The respective feature embeddings are concatenated and projected to the dimension \texttt{embedding\_dim} using a linear layer to produce the content embedding.

The context embedding is computed as the sum of the following feature embeddings:
\begin{enumerate}
    \item Event type and position are encoded using embedding tables with vocabulary sizes of 5 and 256.
    \item Day of week and hour are embedded using tables with vocabulary sizes of 7 and 24.
    \item We also use numeric features: \(\log(1 + \text{hours since last event})\) and \(\log(1 + \text{days since first event})\), each projected to the embedding dimension \texttt{embedding\_dim} via a linear layer.
\end{enumerate}

To encode user history, we use a transformer with causal attention masking. The hidden state corresponding to the final token is used as the user embedding.

\subsubsection{Training Objectives}

We utilize transformer hidden states for various prediction tasks. Specifically, we employ the \textit{Next Event Type Prediction (NETP)} and the \textit{Time Delta Prediction (TDP)}~--- the time gap in \(\log(1 + \text{hours})\) between the current and the next event. For a user sequence of length \(N\) we formulate them as follows:
\[
\mathcal{L}_{\text{NETP}}(\textbf{y}_i, \hat{\textbf{y}_i}) = -\sum_{k=0}^{C-1} y_{i,k} \log \hat{y}_{i,k}
\]
\[
\mathcal{L}_{\text{TDP}}(\boldsymbol{\delta}_i, \hat{\boldsymbol{\delta}}_i) = \left( \delta_{i} - \hat{\delta}_i \right)^2,
\]
where \(C=5\) is the number of event types, \(\textbf{y} \in \mathbb{R}^{(N-1) \times C}\) denotes the one-hot matrix of event types, \(\boldsymbol{\delta} \in \mathbb{R}^{N-1}\) is the time delta between adjacent events, \(\hat{\textbf{y}} \in \mathbb{R}^{(N-1) \times C}\) is the predicted probability distribution over event types, and \(\hat{\boldsymbol{\delta}} \in \mathbb{R}^{N-1}\) denotes the predicted time deltas.  
Note that \(\textbf{y}\) and \(\boldsymbol{\delta}\) represent the targets for the next events in the sequence, and are thus shifted by one position relative to \(\hat{\textbf{y}}\) and \(\hat{\boldsymbol{\delta}}\).

For page visit events, we formulate a \textit{Next URL Prediction (NUP)} as the sampled softmax loss with 8192 in-batch negatives \(\mathcal{N}\):
\[
\mathcal{L}_{\text{NUP}}(u, p) = -\log \frac{e^{\langle g(u), h(p) \rangle / \tau}}{e^{\langle g(u), h(p) \rangle / \tau} + \sum_{i \in \mathcal{N}} e^{\langle g(u), h(d_i) \rangle / \tau}},
\]
where \(\langle g(u), h(d) \rangle / \tau\) is the cosine similarity between normalized user and target embeddings, \(g(u)\) and \(h(d)\), scaled by a learnable temperature parameter \(\tau\).

If the next event is item-specific (i.e., cart addition, purchase, or cart removal), we apply a \textit{Next Item Prediction (NIP)} loss. It is formulated using the sampled softmax loss with the improved logQ correction, which accounts for the fact that the positive item is not Monte Carlo sampled\,\cite{khrylchenko2025correcting}:

\[
\mathcal{L}_{\text{NIP}}(u, p) = - w_{up} \log \frac{e^{\langle g(u), h(p) \rangle / \tau}}{\sum_{i \in \mathcal{N}}^n e^{\langle g(u), h(d_i) \rangle / \tau - \log Q(d_i)}},
\]
where \(w_{up} = \text{sg}(1 - P(p \mid u))\) represents the model's uncertainty\,\cite{khrylchenko2025correcting}, \(\text{sg}\) denotes the stop-gradient operation. Following \citet{yang2020mixed}, \(\mathcal{N}\) is a mixture of 8192 in-batch sampled and 8192 uniformly sampled negatives.

The final loss is obtained as follows:
\begin{align*}
    \mathcal{L}_{\text{SeqEnc}} &=
    \frac{1}{N-1} \sum_{i=0}^{N-2}\mathcal{L}_{\text{NETP}}(\textbf{y}_i, \hat{\textbf{y}}_i)
    + \frac{1}{N-1} \sum_{i=0}^{N-2}\mathcal{L}_{\text{TDP}}(\boldsymbol{\delta}_i, \hat{\boldsymbol{\delta}}_i) \\
    &+ \frac{1}{|\mathcal{I}|} \sum_{(u, p) \in \mathcal{I}} \mathcal{L}_{\text{NIP}}(u, p)
    + \frac{1}{|\mathcal{U}|} \sum_{(u, p) \in \mathcal{U}}\mathcal{L}_{\text{NUP}}(u, p),
\end{align*}
where \(\mathcal{I}\) and \(\mathcal{U}\) are the sets of query-target pairs: \(u\) is the user representation corresponding to the event preceding the item or URL target \(p\), respectively.

Our training procedure is illustrated in Figure~\ref{fig:seq_enc}.

\subsubsection{Implementation Details}
We use \(N=256\) as the maximum sequence length and set \texttt{hidden\_dim=1024, num\_heads=hidden\_dim // 64, dropout=0.1, attn\_dropout=0.1, num\_layers=20}, and use GELU as an activation function. The dimensionality of all embedding tables is set to \texttt{embedding\_dim=256} and the model is trained for 10000 steps with an effective batch size of 8192, which corresponds to approximately five epochs. We use Adam optimizer with a linear learning rate schedule: the learning rate starts at \texttt{1e-5}, linearly increases to \texttt{1e-4} over 3000 warmup steps, and remains constant for the rest of the training. Gradient norms are clipped at 1.0.

\subsection{TwHIN}
Graph neural networks are a popular approach for learning embeddings for both users and items\,\cite{hamilton2020graph}. We have chosen TwHIN\,\cite{el2022twhin} due to its simple parametrization and ability to handle relations (edges) of different types. 

We train the TwHIN model on a bipartite user-item graph with two edge types corresponding to cart additions and purchases.

\subsubsection{Training Objective}
TwHIN is trained on a set of user-item pairs \(E = \{(u_i, d_i)\}_{i=1}^{|E|}\), where \(|E|\) is the total number of edges. We use the \textit{link prediction objective}, formulated as the binary cross-entropy loss. For each edge, we compute the probability logit as the dot product between the normalized user embedding and the normalized sum of the item and edge embeddings:
\[
\mathcal{L}_{\text{TwHIN}}(\mathcal{B}) = -\frac{1}{|\mathcal{B}|} \sum_{i \in \mathcal{B}} \left(\log(p(e_{ii})) + \sum_{j \in \mathcal{B}, j\ne i}\log(1 - p(e_{ij}))\right), 
\]
where \(\mathcal{B}\) is a training batch; \(p(e)\) is the predicted probability of an edge \(e\); \(e_{ij}\) denotes the edge between \(u_i\) and \(d_j\).

\subsubsection{Implementation Details}
The training set includes all users from the competition evaluation set, along with the most active users (based on cart additions and purchases) from the remaining pool, totaling 1.5 million users. Each user is assigned a trainable embedding of size 256. Increasing embedding dimensionality to 512 did not yield improvements.

For items, we use the same encoder architecture as in Section~\ref{sec:sequential}; moreover, we initialize it using the same weights trained during transformer-based sequential modeling. Effective batch size is 4096, learning rate is constant and equal to \texttt{1e-3}. To mitigate overfitting, we apply dropout with probability 0.3 to both user and item embeddings before normalizing them. Gradient norms are clipped at 1.0.

\subsection{Handcrafted Features}
We derive over 300 handcrafted features from the user activity history during the training period. These features serve both as a part of the Universal Behavioral Profile and as features for the deep cross network, which we discuss later. Empirically, we observe that handcrafted features play a crucial role in achieving strong performance.

We organize them into two main categories:

\begin{enumerate}
    \item \textit{Basic features}: interaction counters, aggregate statistics, cumulative price-related features, temporal and frequency descriptors, behavioral ratios, price-based characteristics, and measures of diversity in user interactions. To capture recency effects, each basic feature is calculated over five temporal windows: the entire training period, as well as the previous 60, 28, 14, and 7 days.
    \item \textit{Cluster-based features} are derived by aggregating user interactions within clusters formed from item name embeddings and price segments. Furthermore, we apply exponential weighting to prioritize recent events.
\end{enumerate}

We discuss each feature in Appendix~\ref{sec::appendix}.

\subsection{Deep Cross Network}

\paragraph{Model Inputs}
Sequential user embeddings are first projected to dimensionality 64 via a projection module consisting of two linear layers with an intermediate RMSNorm and a hidden dimension of 128. Handcrafted numeric features are encoded using piecewise linear encoding\,\cite{gorishniy2022embeddings}: each feature's value range is divided into 64 intervals based on 63 quantiles (\(1/64, 2/64, \ldots, 63/64\)), and each interval is further sub-divided into 32 uniform sub-intervals. Each value is mapped to an interval ID, which is then used to lookup an embedding of dimension 2. Numeric feature embeddings are concatenated with sequential user embedding, and the resulting vector is used as the input to the DCN-v2\,\cite{wang2021dcn} model.

\paragraph{Architecture}
The DCN-v2 model consists of three layers, with the transformation at each layer given by:
\[
\mathbf{x}_{l+1} = \mathbf{x}_0 \odot \left( U_l \left( V_l^\top \mathbf{x}_l \right) + \mathbf{b}_l \right) + \mathbf{x}_l,
\]
where \(V_l, U_l \in \mathbb{R}^{d \times r_l}\), \(d\) is the dimension of concatenated numeric and sequential embeddings. The DCN-v2 output is formed by concatenating \(V_l^\top \mathbf{x}_l\) from each layer and is passed through a four-layer DenseNet\,\cite{huang2017densely} to further increase model capacity.

\paragraph{Training Objectives}
The deep cross network is trained on several downstream objectives: binary cross-entropy for user \textit{churn prediction}, and \textit{propensity prediction} of the sets of relevant SKU and category IDs. We also employ \textit{Contrastive Learning (CL)} objective: for each user, the sequential history is augmented twice by randomly masking 30\% of events, then passed through the sequential encoder. The resulting embeddings \(g(u^{\prime})\) and \(g(u^{\prime \prime})\) are processed through the deep cross network, and serve as positive pairs \((p_u^{\prime}, p_u^{\prime \prime})\) for the sampled softmax loss with 8192 in-batch negatives:
\[
\mathcal{L}_{\text{CL}}(p_u^{\prime}, p_u^{\prime \prime}) = -\log \frac{e^{\langle p_u^{\prime}, p_u^{\prime \prime} \rangle / \tau}}{e^{\langle p_u^{\prime}, p_u^{\prime \prime} \rangle / \tau}+ \sum_{n \in \mathcal{N}} e^{\langle p_u^{\prime}, p_n^{\prime \prime} \rangle / \tau}},
\]
where \(\mathcal{N}\) is a set of in-batch negative examples, and \(\tau\) is a learnable temperature parameter.

Targets for churn, SKU propensity, and category propensity are derived from user behavior in the final two weeks of the training period. To avoid label leakage, both the sequential encoder and handcrafted features use only data from earlier periods.

\paragraph{Implementation Details}
We set \(r_0=64\), and \(r_1=r_2=32\). The DCN-v2 output, used as a deep cross network embedding, has dimension 128. DenseNet hidden dimension is 64. The deep cross network is trained for 25000 steps, corresponding to approximately 30 epochs, with an effective batch size of 16384. We use Adam optimizer and assign different fixed learning rates to distinct parameter groups: \texttt{2e-4} for numeric feature embeddings, \texttt{5e-4} for DCN-v2 parameters, and \texttt{1e-3} for all other parameters. Gradient norms are clipped at 1.0.
\section{Ablation Study}

Server-side evaluation is performed on a set of 1 million users, referred to as relevant users. For the purpose of ablation studies, we perform a timestamp-based data split by reserving the most recent four weeks of logs for downstream model training and evaluation. We evaluate performance on a subset of relevant users, each of whom has at least one interaction during the training period. All subsequent results are computed using the evaluation code released by the organizers\footnote{https://github.com/Synerise/recsys2025}. For churn prediction, we report AUC-ROC. For category prediction and SKU prediction tasks, following official evaluation, we use a weighted combination of AUC-ROC and novelty- and diversity-based measures. \textit{All reported results are averaged over 5 runs.}

Regarding the sequential encoder, Table~\ref{tab:encoder_objectives} shows that NIP is the most critical training objective and that all objectives contribute meaningfully. Table~\ref{tab:encoder_size} presents an ablation on the transformer encoder size. Scaling the number of parameters generally improves performance across tasks.

\begin{table}
    \caption{Sequential encoder objectives}
    \label{tab:encoder_objectives}
    \begin{tabular}{l|ccc}
        \toprule
        Objectives & Churn & Category & SKU \\
        \midrule
        All objectives & \textbf{0.7404} & \textbf{0.8048} & 0.7735 \\
        W/o NETP & 0.7379 & \underline{0.8017} & \underline{0.7738} \\
        W/o TDP & 0.7369 & 0.8016 & \textbf{0.7768} \\
        W/o NUP & \underline{0.7391} & 0.8008 & 0.7727 \\
        W/o NIP & 0.7374 & 0.7974 & 0.7695 \\
        \bottomrule
    \end{tabular}
    \Description{Table presents an ablation study for sequential encoder objectives. Eliminating any objective results in metric decrease.}
\end{table}

\begin{table}
    \caption{Sequential encoder size}
    \label{tab:encoder_size}
    \begin{tabular}{ccc|ccc}
        \toprule
        \#params & \texttt{\#layers} & \texttt{hidden\_dim} & Churn & Category & SKU\\
        \midrule
        252M & 20 & 1024 & \textbf{0.7404} & \textbf{0.8048} & 0.7735 \\
        126M & 10 & 1024 & \underline{0.7397} & \underline{0.8030} & \underline{0.7757} \\
        19M  & 6  & 512  & 0.7371 & 0.8013 & \textbf{0.7761} \\
        3M   & 4  & 256  & 0.7329 & 0.7957 & 0.7712 \\
        \bottomrule
    \end{tabular}
    \Description{Table presents an ablation study of Transformer encoder size. Best metrics are achieved on the largest evaluated encoder.}
\end{table}

For TwHIN, we present an ablation study investigating both the graph construction and the choice of item encoder in Table~\ref{tab:twhin}. We compare the use of the pretrained item encoder to a simpler alternative in which the top 1 million most popular items have ID-based embeddings, while all remaining items are mapped to a single shared embedding. Incorporating cart additions as edges significantly improves performance, while using the pretrained item encoder further enhances category and SKU propensity prediction.

\begin{table}
    \caption{Graph structure and TwHIN item encoder. 'P' indicates the purchase graph, 'A' the cart addition graph, and 'Enc.' the use of the pretrained item encoder.}
    \label{tab:twhin}
    \begin{tabular}{l|ccc}
        \toprule
        & Churn & Category & SKU \\
        \midrule
        Graph: P & 0.6083 & 0.6719 & 0.6321 \\
        Graph: P\&A & \textbf{0.6625} & \underline{0.7048} & \underline{0.6719} \\
        Graph: P\&A; Enc. & \underline{0.6401} & \textbf{0.7094} & \textbf{0.6815} \\
        \bottomrule
    \end{tabular}
    \Description{Table presents metrics for TwHIN trained with/without the item encoder and with varying graph structures. Incorporating both cart additions and a pretrained item encoder improves overall performance.}
\end{table}

Contrastive loss for the deep cross network improves performance across all tasks. The corresponding ablation is shown in Table~\ref{tab:contrastive}.

\begin{table}
    \caption{Contrastive loss for deep cross network training}
    \label{tab:contrastive}
    \begin{tabular}{l|ccc}
        \toprule
        Objectives & Churn & Category & SKU \\
        \midrule
        All objectives & \textbf{0.7509} & \textbf{0.8115} & \textbf{0.7952} \\
        W/o CL & \underline{0.7496} & \underline{0.7972} & \underline{0.7862} \\
        \bottomrule
    \end{tabular}
    \Description{This table presents an ablation study showing the impact of contrastive loss on Deep Cross Network performance. Excluding contrastive loss results in degraded performance for all metrics.}
\end{table}

Recall that our Universal Behavioral Profile consists of four concatenated components: 
\begin{enumerate}
    \item Sequential embedding
    \item TwHIN embedding
    \item Deep cross network embedding
    \item Standardized vector of numeric features
\end{enumerate}

Table~\ref{tab:ablation} presents an ablation study of each component. The impact of the TwHIN is relatively low due to overlapping collaborative signals with the sequential embedding, though TwHIN provided slight improvements on the leaderboard. Removing numeric features has little effect, as deep cross network embeddings already incorporate those signals.

Performance on the conversion (hidden1) and price propensity (hidden3) tasks matched that of the open tasks. However, the product propensity task (hidden2), which evaluated performance on products unseen during embedding training, was prone to overfitting.

\begin{table}
    \caption{Universal Behavioral Profile components}
    \label{tab:ablation}
    \begin{tabular}{l|ccc}
        \toprule
        Components & Churn & Category & SKU \\
        \midrule
        All components & \textbf{0.7531} & \underline{0.8147} & \underline{0.7899} \\
        W/o sequential encoder & 0.7502 & 0.8027 & 0.7883 \\
        W/o TwHIN & \underline{0.7530} & 0.8146 & \textbf{0.7901} \\
        W/o deep cross network & 0.7515 & 0.8111 & 0.7849 \\
        W/o numeric features & 0.7523 & \textbf{0.8154} & 0.7889 \\
        \bottomrule
    \end{tabular}
    \Description{Table presents an ablation study for each of four Universal Behavioral Profile components.}
\end{table}
\section{Conclusion}
This paper presents the 4th-place solution by team \texttt{ambitious} in the RecSys Challenge 2025. Our method builds a Universal Behavioral Profile by integrating: (1) a sequential encoder; (2) the TwHIN graph neural network; (3) a deep cross network based on the DCN-v2 architecture; and (4) a rich set of handcrafted features. Ablation studies confirm the importance of the training objectives and solution components.

\bibliographystyle{ACM-Reference-Format}
\bibliography{sample-base}

\balance
\clearpage
\appendix
\onecolumn
\section{Numeric Features}
\label{sec::appendix}

\subsection{Primary Features}

The primary set of features is presented in Table \ref{table:features}. These features were computed over several temporal windows to capture changes in user behavior across different time horizons:

\begin{itemize}
  \item \textbf{7 days} --- reflects short-term activity and the user's current interests.
  \item \textbf{14 days} --- helps to identify medium-term behavioral trends.
  \item \textbf{28 days} --- corresponds to the downstream model training and evaluation period and serves as the main window for calculations.
  \item \textbf{60 days} --- allows for analysis of long-term behavioral patterns.
  \item \textbf{all} --- aggregates computed over the entire available user history.
\end{itemize}

As a result, a total of 205 features were generated by aggregating across different windows.

In addition to the primary features, we developed cluster-based features that capture the user’s purchase history, enabling the model to account for both topical preferences and the dynamics of purchasing behavior.

\subsection{Topical Features}

To capture user interests across topical item groups, we utilized embeddings of item names for purchased items. These embeddings were grouped using the k-means clustering algorithm:
\begin{itemize}
  \item For the \texttt{c5} features, embeddings were clustered into 5 groups, forming features \texttt{c5\_0}, \texttt{c5\_1}, \ldots, \texttt{c5\_4};
  \item For the \texttt{c10} features, embeddings were clustered into 10 groups, forming features \texttt{c10\_0}, \texttt{c10\_1}, \ldots, \texttt{c10\_9}.
\end{itemize}
For each purchase, binary indicators were generated to denote whether a purchase belonged to the corresponding cluster. This approach allowed aggregation of both the count and value of purchases within topical clusters, thereby providing the model with information on user preferences by item type.

\subsection{Features by Price Segments.}

For each purchase, the item was assigned to one of the price segments: \texttt{super\_lo\_price}, \texttt{lo\_price}, \texttt{mid\_lo\_price}, \texttt{mid\_price}, \texttt{mid\_hi\_price}, \texttt{hi\_price}, or \texttt{super\_hi\_price}.  
These binary features enabled aggregation by price segment and analysis of the user’s propensity to purchase in low, medium, or high price categories.

\subsection{Exponentially Weighted Windows (EW-windows)}
To account for the recency of events, each event was assigned a weight that decays exponentially with the time elapsed between the event and the most recent event in the user’s history. The event weight $w$ was computed as follows:
\[
w = \exp\left(-\frac{\Delta t}{\tau \cdot 86400}\right),
\]
where:
\begin{itemize}
  \item $\Delta t$ is the time difference (in seconds) between the event and the last recorded event in the user's history;
  \item $\tau$ is the time decay scale (set to 28, 50, or 100 days);
  \item $86400$ is the number of seconds in a day.
\end{itemize}

The use of exponentially weighted windows enabled more recent purchases to have a greater impact, reduced the influence of outdated events, and made features more sensitive to current user behavior. This approach allowed us to construct various aggregates, for example:
\begin{itemize}
  \item EW-summed purchases within the window (\texttt{ew\_rows}),
  \item EW-sums by cluster (\texttt{ew\_c5\_0}, \ldots, \texttt{ew\_c5\_4}),
  \item EW-sums by price segment (\texttt{ew\_mid\_hi\_price}, etc.),
  \item Shares and ratios within EW-windows (e.g., \texttt{mid\_hi\_lo\_ratio}),
  \item Pricing statistics: mean, maximum, minimum, price range, and coefficient of variation.
\end{itemize}

Such a combination of cluster-based and price-segmented features with exponential weighting enabled the model to accurately capture both the topical and dynamic aspects of user interests.

In total, computation over all windows and groupings resulted in 141 features. These features are discussed in Table \ref{table:buy_features_full}.

\begin{table*}[!ht]
\caption{Basic Features}
\label{table:features}
\centering
\begin{tabular}{p{0.28\textwidth} p{0.67\textwidth}}
\toprule
\textbf{Feature Name} & \textbf{Description} \\
\midrule
\multicolumn{2}{l}{\textbf{Basic Quantitative Features}} \\
add\_all & Total number of cart additions over the entire observation period. \\
rem\_all & Total number of cart removals over the entire observation period. \\
buy\_all & Total number of purchases made by the user. \\
total\_event\_cnt & Aggregate number of all item-related events (additions, removals, purchases). \\
search\_cnt & Number of search queries issued by the user. Reflects the user’s interest in catalog exploration and activity in seeking items. \\
uniq\_buy\_days & Number of unique calendar days on which the user made purchases. \\
\midrule
\multicolumn{2}{l}{\textbf{Average and Aggregated Price Features}} \\
avg\_price\_add & Average price of items added to the cart by the user. \\
avg\_price\_rem & Average price of items removed by the user from the cart. \\
avg\_price\_bucket\_buy & Average price of purchased items. \\
max\_price\_buy & Maximum price of a single item purchased by the user. \\
min\_price\_buy & Minimum price of a purchased item. \\
price\_spread\_buy & Difference between max\_price\_buy and min\_price\_buy. \\
\midrule
\multicolumn{2}{l}{\textbf{Cumulative Price Features}} \\
add\_price\_sum & Total price of all items added to the cart by the user. \\
buy\_price\_sum & Sum of prices of all purchased items. \\
buy\_volume\_hi & Total value of purchases for high-priced items (above the 80th price percentile). \\
buy\_volume\_lo & Total value of purchases for low-priced items (below the 20th price percentile). \\
rem\_price\_sum & Sum of prices of all items removed from the cart. \\
page\_visit\_cnt & Total number of pages visited by the user. \\
page\_per\_buy & Average number of pages viewed per purchase. \\
uniq\_url\_cnt & Number of unique URLs visited by the user. \\
\midrule
\multicolumn{2}{l}{\textbf{Temporal and Frequency Features}} \\
duration\_days & Number of days between the user’s first and last purchase in the history. \\
buy\_freq\_per\_day & Average number of purchases per day during the active period. \\
cart\_action\_freq\_per\_day & Average number of cart actions (additions and removals) per day. \\
\midrule
\multicolumn{2}{l}{\textbf{Behavioral Ratios}} \\
cart\_ctr & Ratio of the number of items added to cart to the total number of events (total\_event\_cnt). \\
remove\_rate & Proportion of cart removals relative to additions. \\
conv\_rate & Proportion of purchases to additions (buy/add). \\
net\_cart\_ratio & Ratio of the difference between additions and removals (add - rem) to the number of additions. \\
cart\_repeat\_rate & Ratio of removals from cart to additions. \\
search\_to\_add & Average number of additions per search query. Reflects the efficiency of search in generating interest. \\
search\_to\_buy & Average number of purchases per search query. Characterizes the efficiency of search in leading to orders. \\
\midrule
\multicolumn{2}{l}{\textbf{Features Related to Purchases of Differently Priced Items}} \\
hi\_price\_buy\_cnt & Number of purchases of items priced above the 80th percentile of all purchase prices. Indicates user’s tendency to buy expensive items. \\
lo\_price\_buy\_cnt & Number of purchases of items below the 20th percentile of purchase prices. Characterizes low-price segment purchases. \\
hi\_price\_share & Share of expensive item purchases among all user’s purchases. \\
lo\_price\_share & Share of inexpensive item purchases among all user’s purchases. \\
buy\_volume\_hi\_share & Share of total amount spent on expensive items in the total purchase sum. \\
buy\_volume\_lo\_share & Share of total amount spent on inexpensive items in the total purchase sum. \\
lo\_hi\_buy\_ratio & Ratio of the number of inexpensive to expensive item purchases. \\
\midrule
\multicolumn{2}{l}{\textbf{Purchase Diversity}} \\
uniq\_sku\_per\_buy & Average number of unique products (SKUs) per purchase. Evaluates order diversity. \\
uniq\_sku\_per\_day & Average number of unique products bought by the user per day. \\
uniq\_cat\_per\_buy & Average number of unique product categories per purchase. Characterizes the breadth of user interests for each purchase. \\
uniq\_cat\_per\_day & Average number of unique categories bought by the user per day. \\
\bottomrule
\end{tabular}
\end{table*}

\begin{table*}[!ht]
\caption{Cluster-based Features}
\label{table:buy_features_full}
\centering
\begin{tabular}{p{0.38\textwidth} p{0.60\textwidth}}
\toprule
\textbf{Feature Name} & \textbf{Description} \\
\midrule
\multicolumn{2}{l}{\textbf{1. Exponentially Weighted Aggregates over Events (EW-windows)}} \\
ew\_rows\_\{28d,50d,100d\} & EW-sum of all purchases within the window. \\
ew\_c5\_\{0..4\}\_\{28d,50d,100d\} & EW-summed purchases for each c5 cluster within the window. \\
ew\_mid\_hi\_price\_\{28d,50d,100d\}, ew\_mid\_price\_\{28d,50d,100d\}, ew\_mid\_lo\_price\_\{28d,50d,100d\} & EW-summed purchases in high, medium, and low price segments, respectively. \\
\midrule
\multicolumn{2}{l}{\textbf{2. Shares and Ratios within EW-windows}} \\
share\_c5\_\{0..4\}\_\{28d,50d,100d\} & Share of purchases for each c5 cluster in the EW-sum. \\
share\_mid\_hi\_price\_\{28d,50d,100d\}, share\_mid\_price\_\{28d,50d,100d\}, share\_mid\_lo\_price\_\{28d,50d,100d\} & Shares of expensive, medium, and inexpensive purchases. \\
mid\_hi\_lo\_ratio\_\{28d,50d,100d\} & Ratio of the share of expensive to inexpensive purchases in the EW-window. \\
\midrule
\multicolumn{2}{l}{\textbf{3. Price Statistics in EW-windows}} \\
ew\_mean\_price\_\{28d,50d,100d\} & EW-average purchase price. \\
ew\_max\_price\_\{28d,50d,100d\}, ew\_min\_price\_\{28d,50d,100d\} & EW-maximum and minimum purchase prices. \\
price\_range\_\{28d,50d,100d\} & Price range within the window. \\
cv\_price\_\{28d,50d,100d\} & Coefficient of price variation within the window. \\
\midrule
\multicolumn{2}{l}{\textbf{4. EW-based Features for Top-SKUs and Top-Categories}} \\
sku\_cnt\_sum\_\{28d,50d,100d\}, sku\_score\_sum\_\{28d,50d,100d\} & EW-summed purchases and EW-weighted sums for top-100 SKUs. \\
cat\_cnt\_sum\_\{28d,50d,100d\}, cat\_score\_sum\_\{28d,50d,100d\} & EW-summed purchases and EW-weighted sums for top-100 categories. \\
\midrule
\multicolumn{2}{l}{\textbf{5. Aggregates across the Entire Purchase History}} \\
rows\_total & Total number of purchases. \\
sum\_price\_total & Total cost of all purchases. \\
mean\_price\_total, median\_price\_total & Mean and median purchase prices. \\
q25\_price\_total, q75\_price\_total & 25th and 75th price percentiles. \\
max\_price\_total, min\_price\_total & Maximum and minimum purchase prices. \\
std\_price\_total & Standard deviation of purchase prices. \\
active\_days\_total & Number of unique active purchasing days. \\
price\_range\_total, iqr\_price\_total & Price range and interquartile range. \\
cv\_price\_total & Coefficient of price variation. \\
mean\_median\_ratio\_total & Ratio of the mean price to the median. \\
mid\_hi\_share\_total, mid\_lo\_share\_total & Shares of expensive and inexpensive purchases. \\
mid\_hi\_lo\_ratio\_total & Ratio of the share of expensive to inexpensive purchases. \\
avg\_tx\_per\_active\_day\_total & Average number of transactions per active day. \\
mid\_hi\_mean\_ratio\_total, mid\_hi\_iqr\_ratio\_total & Ratios of the share of expensive purchases to mean and IQR. \\
\midrule
\multicolumn{2}{l}{\textbf{6. Features by Price Segment (PRICE7)}} \\
super\_lo\_price\_total, lo\_price\_total, & \\ mid\_lo\_price\_total, mid\_price\_total, & Sums of purchases within each of the 7 price segments. \\ mid\_hi\_price\_total, hi\_price\_total, & \\ super\_hi\_price\_total & \\
\midrule
\multicolumn{2}{l}{\textbf{7. Features by Product Cluster (c5)}} \\
c5\_\{0..4\}\_total & Sums of purchases within each c5 cluster. \\
\midrule
\multicolumn{2}{l}{\textbf{8. Recency Metrics}} \\
gap\_ew\_50d & EW-average time between purchases within the 50-day window. \\
\bottomrule
\end{tabular}
\end{table*}

\clearpage
\section{Solution Overview}
\label{sec::appendix_solution}

\begin{figure*}[]
    \centering
    \includegraphics[width=0.85\textwidth]{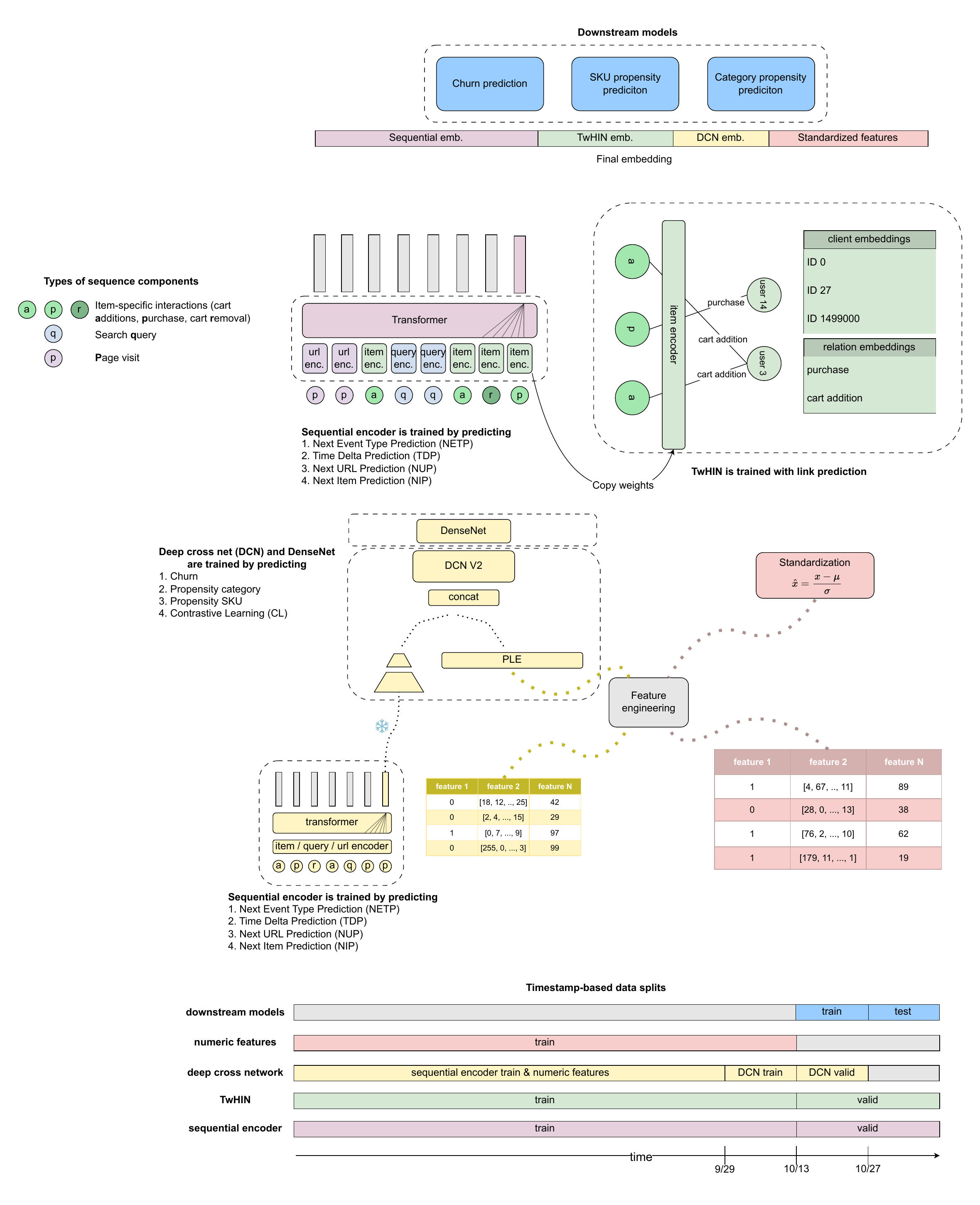}
    \caption{Solution Overview.}
    \Description{}
    \label{fig:solution}
\end{figure*}

Figure~\ref{fig:solution} provides an overview of the four components of our solution. Note that the two sequential encoders are trained on distinct time periods and therefore do not share weights. The timestamp-based data splits are depicted at the bottom of the figure.

\end{document}